\documentclass[aps,prl,showpacs,superscriptaddress,amsmath,amssymb,twocolumn]%
{revtex4}
\usepackage{graphicx}
\usepackage{times}
\usepackage[T1]{fontenc}
\usepackage[svgnames]{xcolor}

\newcommand{\heading}[1]{\par\textbf{#1}\quad\ignorespaces}
\newcommand{\Eq}[1]{Eq.~(\ref{eq:#1})}
\newcommand{\D}{\mathrm{d}}\newcommand{\I}{\mathrm{i}}
\newcommand{\Exp}[1]{\mathrm{e}^{\mbox{\footnotesize$#1$}}}
\newcommand{\tr}[1]{\textnormal{tr}{\left\{#1\right\}}}
\newcommand{\intr}{\int(\D\vec{r})\,}
\newcommand{\intrp}{\int\!\frac{(\D\vec{r})(\D\vec{p})}{(2\pi\hbar)^2}}

\DeclareMathAlphabet{\vecfont}{OT1}{cmr}{bx}{it}
\renewcommand{\vec}[1]{\vecfont{#1}}
\newcommand{\grad}{\boldsymbol{\nabla}}  
\begin{document}
\title{Leading gradient correction to the kinetic energy for two-dimensional
  fermion gases}
\author{Martin-Isbj\"orn Trappe}
\affiliation{Centre for Quantum Technologies, National University of
  Singapore, 3 Science Drive 2, Singapore 117543, Singapore} 
\author{Yink Loong Len}
\affiliation{Centre for Quantum Technologies, National University of
  Singapore, 3 Science Drive 2, Singapore 117543, Singapore} 
\affiliation{Data Storage Institute, Agency for Science, Technology, and
  Research,\\ 5 Engineering Drive I, Singapore 117608, Singapore} 
\affiliation{Department of Physics, National University of Singapore, 2
  Science Drive 3, Singapore 117542, Singapore} 
\author{Hui Khoon Ng}
\affiliation{Centre for Quantum Technologies, National University of
  Singapore, 3 Science Drive 2, Singapore 117543, Singapore} 
\affiliation{Yale-NUS College, 16 College Avenue West, %
                  Singapore 138527, Singapore}
\affiliation{MajuLab, CNRS-UNS-NUS-NTU International Joint Unit,
UMI 3654, Singapore}
\author{Cord Axel M\"uller}
\affiliation{Centre for Quantum Technologies, National University of
  Singapore, 3 Science Drive 2, Singapore 117543, Singapore} 
\affiliation{Fachbereich Physik, Universit\"at Konstanz, %
  78457 Konstanz, Germany}
\author{Berthold-Georg Englert}
\affiliation{Centre for Quantum Technologies, National University of
  Singapore, 3 Science Drive 2, Singapore 117543, Singapore} 
\affiliation{Department of Physics, National University of Singapore, 2
  Science Drive 3, Singapore 117542, Singapore} 
\affiliation{MajuLab, CNRS-UNS-NUS-NTU International Joint Unit,
UMI 3654, Singapore}
 
\date{25 February 2016}

\begin{abstract}
Density functional theory (DFT) is notorious for the absence of gradient
corrections to the two-dimensional (2D) Thomas-Fermi kinetic-energy functional;
it is widely accepted that the 2D analog of the 3D von
Weizs\"acker correction vanishes, together with all higher-order corrections.
Contrary to this long-held belief, we show that the leading correction to the
kinetic energy does not vanish, is unambiguous, and contributes perturbatively
to the total energy.
This insight emerges naturally in a simple extension of standard DFT, which has
the effective potential energy as a functional variable on equal footing with
the single-particle density. 
\end{abstract}


\begin{widetext}
\maketitle  
\end{widetext}

\heading{Introduction}
Recent advances in the experimental creation and control of
ultracold Fermi gases 
\cite{Giorgini+2:08} 
in two-dimensional (2D) geometries \cite{Martiyanov2010,Dyke2011,Makhalov2014,Ries2015,Fenech2016,Boettcher2016} 
have triggered theoretical work on the semiclassical description 
of fermionic atoms by density functional theory (DFT)
\cite{Fang+1:11,vanZyl+2:13,vanZyl2014,vanZyl2015}.
While the Thomas-Fermi (TF) approximation to the kinetic-energy functional 
is accurate enough at the early stage of these investigations, 
better approximations will eventually be required for, say,
more precise thermometry 
\cite{Stewart2006,Lu2012,Aikawa+5:14} and 
more realistic descriptions of interfaces in multi-component Fermi gases  
\cite{Partridge2006,Du2008,Ketterle2009,Conduit+1:09,Zwierlein2011,Sanner+5:12,TrGrBrRz2015}.
The DFT formalism of Kohn-Sham (KS) type handles the kinetic-energy contribution accurately, but at the price of a large overhead of single-particle orbitals \cite{Karlicky2012,Rasmussen2015,Hu2015}. 
Whereas highly precise KS calculations are standard fare in 3D chemical physics and material science, 
%from neutral atoms \cite{Perdew2006,Lee2009} to 2D materials, 
the problematic dimensional reduction to 2D requires tailored approximations of the exchange-correlation functional, for which no general consensus has been reached \cite{Constantin2008,Raesaenen2010b,Chiodo2012}. Generally, significant efforts are spent on improving orbital-free approximations of functionals, not only within the KS scheme \cite{Perdew2006,Lee2009,Sun2012,Sun2013}, and particularly in 2D  \cite{Jiang2004,Pittalis2009,Raesaenen2010}, but foremost because an accurate orbital-free DFT would excel by superior computational efficiency \cite{Cangi2011,Burke2012,Karasiev2013,Burke2015}.

Improving upon the TF kinetic energy functional requires gradient terms that account for the inhomogeneity in the
single-particle density to leading order.
Unfortunately, at first sight it appears that the 2D analog of the 3D von Weizs\"acker (vW)
correction has a vanishing coefficient, and that all
higher-order corrections vanish, too.
This has been known for decades, at least since the early 1990s
\cite{Holas+2:91,Shao:93}, and has become generally accepted wisdom (see, for
example, \cite{Brack+1:03,Koivisto+1:07,Putaja+4:12}).
We are thus confronted with a dilemma:
On the one hand, we know that the TF approximation cannot be exact; 
on the other hand, there is no established pathway toward nonzero corrections.
It is understandable, then, that various ad-hoc corrections have been invented, such as the
vW-type term  \cite{vanZyl+2:13} and the nonlocal average-density functional recently proposed by van~Zyl \textit{et al.} \cite{vanZyl2014,vanZyl2015}.

However, systematic progress is possible without improvisation. In this Letter, we provide an analytical, orbital-free approach to the calculation of the leading gradient correction to the TF kinetic-energy functional. By a simple extension of standard DFT, which uses
the effective potential energy as an independent variable on equal footing
with the single-particle density \cite{Englert:92}, we obtain a nonzero
gradient correction that is unambiguous and yields a first-order
correction to the energy that can be evaluated by the usual
perturbation-theory method. 
The problem with, and the ambiguities of, the gradient correction to the
density functional arise when one eliminates the effective potential energy
in order to arrive at a functional of the density alone.

\heading{Functionals}
We review briefly the construction of the joint functional of the
single-particle density $n(\vec{r})$ and the effective potential energy
$V(\vec{r})$, as given in \cite{Englert:92}.
We incorporate the particle-count constraint
\begin{equation}\label{eq:A1}
  N=\intr n(\vec{r})
\end{equation}
into the density functional
\begin{equation}\label{eq:A2}
  E[n]=E_{\mathrm{kin}}[n]+\intr V_{\mathrm{ext}}(\vec{r})n(\vec{r})
       +E_{\mathrm{int}}[n]
\end{equation}
with the aid of a Lagrange multiplier, the chemical potential~$\mu$,
\begin{equation}\label{eq:A3}
    E[n,\mu]=E[n]+\mu{\left(N-\intr n(\vec{r})\right)}\,.
\end{equation}
Here $(\D\vec{r})$ denotes the volume element at position $\vec{r}$;
$V_{\mathrm{ext}}(\vec{r})$ is the external potential energy for a probe
particle at $\vec{r}$; $N$ is the total number of particles;
$E_{\mathrm{kin}}[n]$ is the density functional of the kinetic energy; and
$E_{\mathrm{int}}[n]$ is that of the particle-particle interaction energy.
The response of $E_{\mathrm{kin}}[n]$ to variations of the density identifies
the effective potential energy $V(\vec{r})$,
\begin{equation}\label{eq:A4}
  \delta E_{\mathrm{kin}}[n]=-\intr \delta n(\vec{r})[V(\vec{r})-\mu]\,,
\end{equation}
and the Legendre transformation
\begin{equation}\label{eq:A5}
  E_1[V-\mu]=E_{\mathrm{kin}}[n]+\intr[V(\vec{r})-\mu]n(\vec{r})
\end{equation}
introduces the potential-energy functional $E_1[V-\mu]$, since
\begin{equation}\label{eq:A6}
  \delta E_1[V-\mu]=\intr\delta[V(\vec{r})-\mu]\,n(\vec{r})
\end{equation}
has no contribution associated with $\delta n(\vec{r})$.
Accordingly, we have the joint functional
\begin{eqnarray}\label{eq:A7}
  E[V,n,\mu]&=&E_1[V-\mu]-\intr[V(\vec{r})-V_{\mathrm{ext}}(\vec{r})]n(\vec{r})
\nonumber\\ &&\mbox{}+E_{\mathrm{int}}[n]+\mu N\,,
\end{eqnarray}
which is stationary at the actual $V(\vec{r})$, $n(\vec{r})$, and $\mu$.

The structure of \Eq{A5} shows that ${E_1[V-\mu]}$ is the
expectation value of
\begin{equation}\label{eq:A8}
  \sum_{k=1}^N[T(\vec{P}_k)+V(\vec{R}_k)-\mu]
  =\sum_{k=1}^N[H(\vec{R}_k,\vec{P}_k)-\mu]\,,
\end{equation}
the Hamiltonian of noninteracting particles with kinetic energy $T(\vec{p})$
and potential energy ${V(\vec{r})-\mu}$ for each particle, in the $N$-particle
ground state of the physical Hamiltonian
\begin{equation}\label{eq:A9}
  \sum_{k=1}^N[T(\vec{P}_k)+V_{\mathrm{ext}}(\vec{R}_k)]+H_{\mathrm{int}}
\end{equation}
that involves the potential energy $V_{\mathrm{ext}}(\vec{r})$ of the external
forces and the full $N$-particle interaction Hamiltonian 
$H_{\mathrm{int}}$~\cite{rp-RP}. 

The vanishing linear response of $E[V,n,\mu]$ to variations 
$\delta V(\vec{r})$, $\delta n(\vec{r})$, and $\delta\mu$ implies the set of
equations
\begin{subequations}\label{eq:A10}
\begin{eqnarray}
\label{eq:A10a}
\makebox[5em][s]{${\delta V:}$ $n(\vec{r})$}
&=&\frac{\delta}{\delta V(\vec{r})}E_1[V-\mu]\,,\\
\label{eq:A10b}
\makebox[5em][s]{${\delta n:}$ $V(\vec{r})$}
&=&V_{\mathrm{ext}}(\vec{r})+\frac{\delta}{\delta
  n(\vec{r})}E_{\mathrm{int}}[n]
\,,
%\\
%\label{eq:A10c}
%\makebox[5em][s]{${\delta \mu:}$ $N$}
%&=&\intr n(\vec{r})\,,    
\\
\label{eq:A10c}
\makebox[5em][s]{${\delta \mu:}$ $N$}
&=&-\frac{\partial}{\partial\mu}E_1[V-\mu]\,,  
\end{eqnarray}
\end{subequations}
jointly solved by the actual effective potential energy $V(\vec{r})$, the
actual single-particle density $n(\vec{r})$, and the actual value of the
chemical potential $\mu$. Equation \eqref{eq:A1} is recovered by combining Eqs.~(\ref{eq:A10a}) and (\ref{eq:A10c}).
We can convert $E[V,n,\mu]$ into a functional $E[V,\mu]$ of
$V(\vec{r})$ and $\mu$ by solving \Eq{A10b} for $n(\vec{r})$ in terms of
$V(\vec{r})$. 
Likewise, we return from $E[V,n,\mu]$ to $E[n,\mu]$ by solving \Eq{A10a} for
$V(\vec{r})$ in terms of $n(\vec{r})$ and using this $V(\vec{r})$ in
$E[V,n,\mu]$.
In particular, the kinetic-energy density functional is obtained as
\begin{equation}\label{eq:A11}
  E_{\mathrm{kin}}[n]
=\biggl(\!E_1[V-\mu]-\!\intr[V(\vec{r})-\mu]n(\vec{r})\!\biggr)
\Biggr|
_{\mbox{\footnotesize{$\begin{array}[t]{l}V(\vec{r})-\mu\\
\mbox{from \eqref{eq:A10a}}\end{array}$}}}
\end{equation}
\emph{provided that} we can carry out the necessary steps.
For the familiar TF model for the 3D electron gas in atoms, these matters are
discussed in \cite{Englert:88}. 

As an example in 2D, we consider a gas of $N$ unpolarized spin-$\frac{1}{2}$
atoms of mass $m$ with a repulsive contact interaction of strength $W>0$.
We have
\begin{eqnarray}\label{eq:A12}
  E[V,n,\mu]&=&-\frac{m}{2\pi\hbar^2}\intr[\mu-V(\vec{r})]_+^2\nonumber\\
            &&\mbox{} -\intr[V(\vec{r})-V_{\mathrm{ext}}(\vec{r})]n(\vec{r})
           \nonumber\\&&\mbox{}
             +\frac{W}{2}\intr n(\vec{r})^2 +\mu N
\end{eqnarray}
in TF approximation, where $\vec{r}$ is now a 2D position vector and
$(\D\vec{r})$ is its area element, and $[x]_+$ selects the positive values of
variable $x$, that is: $[x]_+=x\eta(x)$,
with Heaviside's unit step function $\eta(\ )$.
The actual $V(\vec{r})$, $n(\vec{r})$, and $\mu$ solve
\begin{subequations}\label{eq:A14}
\begin{eqnarray}
\label{eq:A14a}  n(\vec{r})&=&\frac{m}{\pi\hbar^2}[\mu-V(\vec{r})]_+\,,\\
\label{eq:A14b}  V(\vec{r})&=& V_{\mathrm{ext}}(\vec{r})+W n(\vec{r})\,,%\\
%\label{eq:A14c} N&=&\intr n(\vec{r})\,,
\end{eqnarray}
\end{subequations}
resulting in 
%\begin{equation}%\label{eq:A15}
$  
n(\vec{r})=(W+\pi\hbar^2/m)^{-1}[\mu-V_{\mathrm{ext}}(\vec{r})]_+
$
%\end{equation}
for the density, 
with the value of $\mu$ determined by \Eq{A1}, and the
effective potential energy then from \Eq{A14b}.

The kinetic-energy functional 
\begin{equation}
  \label{eq:A16}
  E_{\mathrm{kin}}[n]=\frac{\pi\hbar^2}{2m}\intr n(\vec{r})^2
\end{equation}
is obtained in accordance with \Eq{A11}, and we note that solving
\Eq{A14a} for ${V(\vec{r})-\mu}$ in terms of $n(\vec{r})$ is only possible
where ${n(\vec{r})>0}$, whereas this equation does not tell us the value of
$V(\vec{r})$ where the density vanishes. 
This is of no consequence in this example, but the proviso at \Eq{A11} must
be kept in mind.

The reduced functionals
\begin{equation}\label{eq:A17}
  E[n]=\frac{\pi\hbar^2+mW}{2m}\intr n(\vec{r})^2
       +\intr V_{\mathrm{ext}}(\vec{r}) n(\vec{r})
\end{equation}
and
\begin{eqnarray}\label{eq:A18}
  E[V,\mu]&=&-\frac{m}{2\pi\hbar^2}\intr[\mu-V(\vec{r})]_+^2
         \\ \nonumber &&\mbox{}
         -\frac{1}{2W}\intr[V(\vec{r})-V_{\mathrm{ext}}(\vec{r})]_+^2
         +\mu N
\end{eqnarray}
are clearly quite different, they are not just reparameterizations of each
other. 
The density functional $E[n]$ is minimal for the actual density whereas the
potential-energy functional is maximal for the actual potential energy and
the actual value of the chemical potential,
\begin{equation}\label{eq:A19}
E(N)=\min_{n}\{E[n]\}=\max_{V,\mu}\{E[V,\mu]\}\,,
\end{equation}
where the permissible densities obey the constraint of \Eq{A1}. 
We get upper bounds on the actual energy $E(N)$ from trial densities in
$E[n]$, and lower bounds from trial values for $V(\vec{r})$ and $\mu$ in
$E[V,\mu]$. 

We must note in this context that the potential functionals of
\cite{Cangi+2:13} are \emph{not} of the $E[V,\mu]$ kind.
Rather, they are density functionals of the usual $E[n]$ kind in disguise,
with the density parameterized in terms of the external potential (as one does
at an intermediate step in the standard proof of the Hohenberg-Kohn theorem).
Since $E[n]$ provides upper bounds on the actual energy, so do these 
functionals of the $E\bigl[n[V_{\mathrm{ext}}]\bigr]$ type.

\heading{Gradient corrections}
The Hamiltonian in \Eq{A8} is that of noninteracting particles, with the
kinetic energy $T(\vec{p})=\vec{p}^2/(2m)$ fixed and different choices for 
$V(\vec{r})-\mu$. 
Since there is one copy of the single-particle Hamiltonian $H(\vec{R},\vec{P})$
for each particle, it follows that $E_1[V-\mu]$ is the trace of
some function $f(H-\mu)$ of $H(\vec{R},\vec{P})-\mu$ \cite{Englert:92}.
For truly noninteracting fermions, all single-particle orbitals with
energies below the chemical potential would be occupied and none above, so
that ${f(x)=x\eta(-x)=-[-x]_+}$ then.
For interacting fermions, this $f(x)$ is an approximation, but it is
sufficiently accurate for the current purpose, and so we approximate
$E_1[V-\mu]$ by 
\begin{eqnarray}\label{eq:B1}
  E_1[V-\mu]&=&-\tr{\bigl[\mu-H(\vec{R},\vec{P})\bigr]_+}\nonumber\\
&=&-2\intrp\Bigl([\mu-H\bigr]_+
\Bigr)_\textsc{w}(\vec{r},\vec{p})\,,\qquad
\end{eqnarray}
where we exhibit a factor of two for the spin multiplicity and evaluate the
quantum-mechanical trace by the phase-space integral of the Wigner function
for the single-particle operator $[\mu-H(\vec{R},\vec{P})]_+$. 

The lowest-order terms in a gradient expansion of the Wigner function
$[g(A)]_\textsc{w}$ of an operator function $g(A)$ in terms of the 
Wigner function $A_\textsc{w}$ of the argument are 
\cite{Wigner-lead-corr}
\begin{eqnarray}\label{eq:B2}
  [g(A)]_\textsc{w}&=&g(A_\textsc{w})
   -\frac{\hbar^2}{16}\bigl\{A_\textsc{w}\Lambda^2A_\textsc{w}\bigr\}
    g''(A_\textsc{w})\nonumber\\&&\mbox{}
   +\frac{\hbar^2}{24}
   \bigl\{A_\textsc{w}\Lambda A_\textsc{w}\Lambda A_\textsc{w}\bigr\}
    g'''(A_\textsc{w})\,,
\end{eqnarray}
where ${\Lambda=\loarrow{\partial_{\vec{r}}}\cdot
 \roarrow{\partial_{\vec{p}}}
  -\loarrow{\partial_{\vec{p}}}\cdot
 \roarrow{\partial_{\vec{r}}}}$
is the two-sided differential operator of the classical Poisson bracket, which
acts only on the $A_\textsc{w}$ factors standing right next to it inside the
curly brackets, and terms of order $(\hbar\Lambda)^4$ and higher are neglected
in \Eq{B2}.
For ${A=H(\vec{R},\vec{P})-\mu}$ with ${A_\textsc{w}=H(\vec{r},\vec{p})-\mu}$ 
and ${g(x)=[-x]_+}$ with the derivatives ${g''(x)=\delta(x)}$ and
${g'''(x)=\delta'(x)}$, we find
\begin{eqnarray}\label{eq:B4}
  E_1[V-\mu]&=&-\frac{m}{2\pi\hbar^2}\intr[\mu-V(\vec{r})]_+^2
             \nonumber\\&&\mbox{}
              +\frac{1}{24\pi}\intr\delta\bigl(\mu-V(\vec{r})\bigr)
              \bigl[\grad V(\vec{r})\bigr]^2.\qquad
\end{eqnarray}
The first term is the TF approximation that was already used in \Eq{A12}, and
the second term --- of second order in the gradient --- is the leading quantum
correction, formally of relative size $\propto\hbar^2$.  
The resulting quantum correction to the TF energy is obtained by a
perturbative evaluation, 
\begin{equation}\label{eq:B5}
  \Delta_{\mathrm{qu}}E=\frac{1}{24\pi}\intr\delta(\mu-V_\textsc{tf})
               (\grad V_\textsc{tf})^2\,,       
\end{equation}
with the effective potential energy $V_{\textsc{tf}}(\vec{r})$ found in the TF
approximation that neglects the second term in \Eq{B4}. 
Exceptional cases aside, the gradient of $V_{\textsc{tf}}$ is continuous at
the border between the classically allowed and forbidden regions selected by
the delta function, and there is no ambiguity in evaluating the integral
\cite{ourEX}.  

In view of \Eq{A11}, this $\Delta_{\mathrm{qu}}E$ is also the quantum
correction that the leading correction to the TF approximation of
$E_{\mathrm{kin}}[n]$ in \Eq{A16} should produce.
We find this corresponding gradient correction by solving
\begin{equation}\label{eq:B6}
  n=\frac{m}{\pi\hbar^2}[\mu-V]_+
     +\frac{1}{24\pi}\bigl[\grad^2\eta(\mu-V)
                          +\delta(\mu-V)\grad^2(\mu-V)\bigr]
\end{equation}
for $\mu-V$ in terms of $n$ up to second order in the gradient,
\begin{equation}\label{eq:B7}
  \mu-V=\frac{\pi\hbar^2}{m}{\left[n-\frac{1}{24\pi}\grad^2\eta(n)
                              -\frac{1}{24\pi}\delta(n)\grad^2n\right]}\,,
\end{equation}
and then using this in \Eq{A11} to arrive at
\begin{equation}\label{eq:B8}
  E_{\mathrm{kin}}[n]=\frac{\pi\hbar^2}{2m}\intr{\left[n(\vec{r})^2
   +\frac{1}{12\pi}\delta\bigl(n(\vec{r})\bigr)
    \bigl(\grad n(\vec{r})\bigr)^2\right]}\,.
\end{equation}

The correction term ${\propto\delta(n)(\grad n)^2}$ is well known \cite{vW},
but not universally established.
It has been found by some methods used for deriving gradient corrections
\cite{KKvsW} (see, for example, \cite{Brack+1:03,vanZyl:00}) or not found 
by other methods (see, for example,
\cite{Holas+2:91,Shao:93,Koivisto+1:07,Putaja+4:12}). 
When the term was found, it was discarded on the basis that it gives ``a
vanishing contribution to the integrated kinetic energy for physical densities
which decay \emph{smoothly} to zero as $r$ tends to infinity''
\cite{vanZyl+2:13}, which is a reasonable argument.

In any case, the correction term is rather problematic.
Recalling the remark after \Eq{A16}, we observe that \Eq{B7} is restricted to
regions where ${n(\vec{r})>0}$, and there we have ${\delta(n)=0}$.
But what about the border region that solely contributes to
$\Delta_{\mathrm{qu}}E$ in \Eq{B5}?
Further, an attempt at a perturbative evaluation,
\begin{equation}\label{eq:B9}
  \Delta_{\mathrm{qu}}E=\frac{\hbar^2}{24m}\intr \delta(n_{\textsc{tf}})
        (\grad n_{\textsc{tf}})^2\,,
\end{equation}
requires the assignment of a value to $(\grad n_{\textsc{tf}})^2$ where the
gradient of $n_{\textsc{tf}}$ is discontinuous.
This is in marked contrast to $\Delta_{\mathrm{qu}}E$ in \Eq{B5} where
$\grad V_{\textsc{tf}}$ is (usually) continuous across the border
between the classically allowed region ($\mu>V$) and the classically forbidden
region ($\mu<V$).  

Clearly, these problems occur in the transition from $E[V,n,\mu]$ to
$E[n,\mu]$ and, eventually, to $E[n]$.
We can stay out of trouble by consistently working with the joint functional
$E[V,n,\mu]$. 
Also in other contexts, functionals of the effective potential energy have
been more useful than the standard functionals of DFT \cite{V-func}.

Not only the correction term ${\propto\delta(n)(\grad n)^2}$ has been found
before, also intermediate equations such as \Eq{B6} or similar appear in other
derivations --- with a colossal difference in physical meaning, however:
The effective potential energy $V(\vec{r})$ is a variable of the
functional $E[V,n,\mu]$ on equal footing with the density $n(\vec{r})$ and we
prefer to keep $V$ in the formalism, rather than eliminating it.
In other derivations, an auxiliary variable $V(\vec{r})$ is introduced as a
technical tool for deriving statements about systems of noninteracting
particles, is eliminated at the earliest convenience without a trace, 
and is never a variable of a functional. 
It is also worth remembering that the effective potential energy accounts for
the interaction fully [see \Eq{A10b}], and the functional $E_1[V-\mu]$, be it in
TF approximation or beyond, is equally valid for interacting and
noninteracting particles.

%\begin{figure}
%\centerline{\includegraphics{GradCorr2D-Fig1v2}}
%\caption{\label{fig:HO}%
%Noninteracting spin-$\frac{1}{2}$ particles in an external harmonic-oscillator
%potential. The plot shows, for $N=1,2,\dots,30$, the difference between the
%exact energy $E_{\mathrm{ex}}(N)$ and its TF approximation
%$E_{\textsc{tf}}(N)$ divided by $\hbar\omega N^{\frac{1}{2}}$.
%This is an oscillatory contribution to the energy, smallest for the
%closed-shell values $N=2,6,12,20,30,\dots$, and largest for half-filled shells.
%The centers of the squares are the exact values, connected by straight
%dashed-line segments that guide the eye.   
%The horizontal line shows 
%$\Delta_{\mathrm{qu}}E(N)/(\hbar\omega N^{\frac{1}{2}})$ of \Eq{C4}.}
%\end{figure}

\begin{figure}
\centerline{\includegraphics[width=0.9\linewidth]{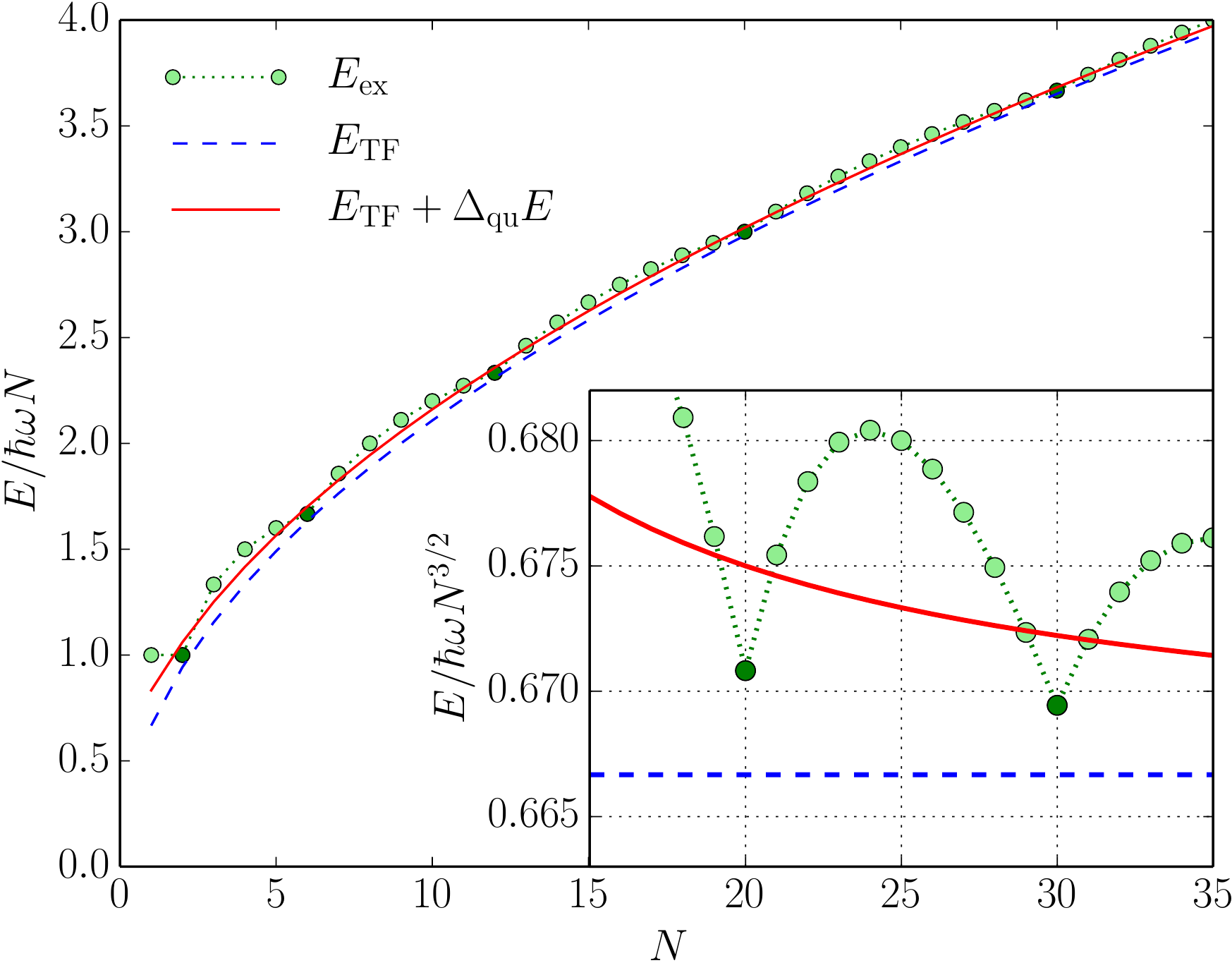}}
\caption{\label{fig:HO}%
Energy of $N$ noninteracting fermions in an isotropic 2D harmonic trap with angular frequency $\omega$. Main plot: energy per particle. Inset: Zoom on the energy divided by the TF scaling $N^{3/2}$. Circles connected by dotted lines are the exact energies \cite{HarmOscEnergy}; dark circles denote the closed-shell values ${N=2,6,12,20,30}$. The dashed blue line is the TF approximation, \Eq{C3}, consistently below the exact energies. The full red line includes the leading quantum correction \Eq{C4}, interpolating the oscillations of the exact values above the TF result.}
\end{figure}

\heading{2D harmonic oscillator}
The external harmonic-oscillator potential ${V_{\mathrm{ext}}(\vec{r})=\frac{1}{2}m\omega^2\vec{r}^2}$ is omnipresent in trapped 2D Fermi gases \cite{Stanescu2007,Martiyanov2010,Dyke2011,Makhalov2014,Ries2015,Fenech2016,Boettcher2016} and often appropriate for other systems, like electrons in quantum dots \cite{Reimann2002}. It is good practice to employ exactly solvable models for judging the accuracy of approximate energy functionals as done in \cite{Holas+2:91,Brack2001,vanZyl+2:13} for harmonically confined noninteracting particles \cite{KSvsSemiclassics}. We follow this tradition and examine $E[V,n,\mu]$ in TF approximation,
\begin{eqnarray}\label{eq:C1}
  E[V,n,\mu]&=&-\frac{m}{2\pi\hbar^2}\intr[\mu-V(\vec{r})]_+^2\\ \nonumber
     &&\mbox{} -\intr\Bigl[V(\vec{r})-\frac{1}{2}m\omega^2\vec{r}^2\Bigr]
                      n(\vec{r})  +\mu N \,. 
\end{eqnarray}
The stationary values are 
$V_{\textsc{tf}}(\vec{r})=V_{\mathrm{ext}}(\vec{r})%
=\frac{1}{2}m\omega^2\vec{r}^2$, of
course, as well as
\begin{equation}\label{eq:C2}
  n_{\textsc{tf}}(\vec{r})=\frac{m\omega}{\pi\hbar}
             \Bigl[N^{\frac{1}{2}}-\frac{m\omega}{2\hbar}\vec{r}^2\Bigr]_+
   \quad\mbox{and}\quad \mu_{\textsc{tf}}=\hbar\omega N^{\frac{1}{2}}\,.
\end{equation}
They yield the TF energy
\begin{equation}\label{eq:C3}
  E_{\textsc{tf}}(N)=-\frac{1}{3}\hbar\omega N^{\frac{3}{2}}
                +0+\hbar\omega N^{\frac{3}{2}}
               =\frac{2}{3}\hbar\omega N^{\frac{3}{2}}\,,
\end{equation}
where the three-term sum refers to the three contributions in \Eq{C1}.
The quantum correction of \Eq{B5} is
\begin{equation}\label{eq:C4}
  \Delta_{\mathrm{qu}}E(N)=\frac{1}{6}\hbar\omega N^{\frac{1}{2}}\,,
\end{equation}
which is unambiguous, definitely nonzero, and small compared with the leading
TF contribution.
Figure~\ref{fig:HO} shows that $\Delta_{\mathrm{qu}}E(N)$ gives an average
account of the oscillatory difference between the exact energy
\cite{HarmOscEnergy} and the TF approximation.

\begin{figure}
\centerline{\includegraphics[width=220\unitlength]{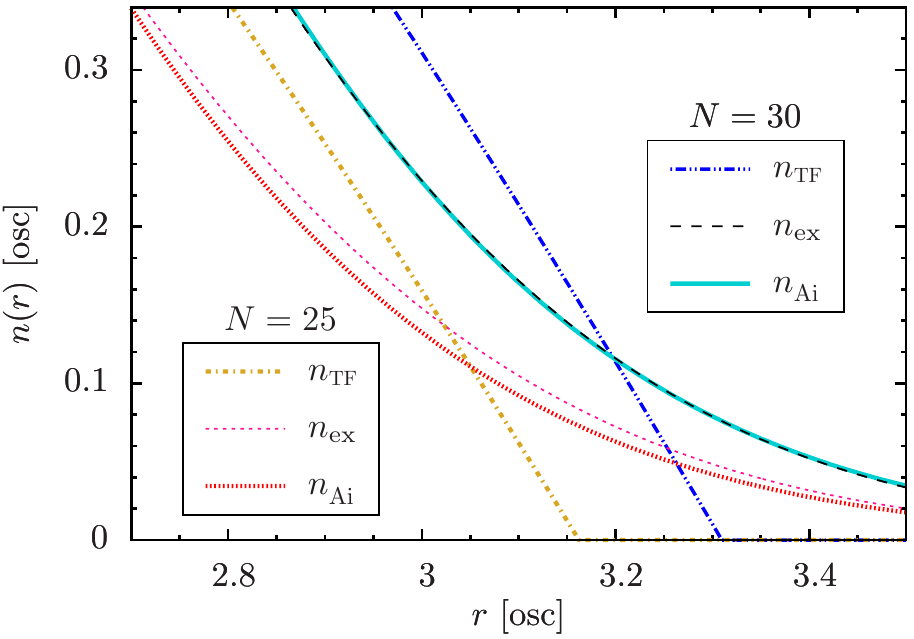}}
\caption{\label{fig:Airy}%
Densities $n(r)$ of noninteracting spin-$\frac{1}{2}$ particles in
an external harmonic-oscillator potential in the vicinity of the border
between the classically allowed and forbidden regions.  
The abscissa is the radial distance $r=|\vec{r}|$ in oscillator units.
For five filled shells (${N=30}$) and for the fifth shell half-full (${N=25}$),
the plot shows the exact densities $n_{\mathrm{ex}}^{\ }$, their TF approximations
$n_{\textsc{tf}}^{\ }$, and the densities $n_{\mathrm{Ai}}^{\ }$ obtained with
Airy-averaging techniques.}
\end{figure}

\heading{Particle density}
The leading gradient correction of \Eq{B4} is fine for the perturbative
evaluation as in \Eq{B5} but the implied correction to the single-particle
density in \Eq{B6} is singular and entirely
localized at the border between the classically allowed and forbidden regions.
A fully satisfactory improvement over the TF approximation should yield a
smooth transition across this border.
This is achieved with the 2D analogs of the 3D Airy-averaging 
techniques \cite{Englert+1:84b,Englert:88}, by which one obtains better
approximations for $E_1[V-\mu]$ and the resulting density~\cite{Airy-av}.
These matters and others will be discussed elsewhere \cite{Trappe+4:IP}.
Here we are content with showing, in Fig.~\ref{fig:Airy}, two such densities
for the harmonic-oscillator example above, together with the exact densities
and their TF approximations.
Clearly, the Airy averages improve matters much and yield very reasonable
densities \cite{Alt-E1}.

\heading{Summary}
We established the leading gradient correction to the TF approximation for the
kinetic energy for a 2D gas of fermions.
This quantum correction is unambiguous and its nonzero contribution to the
energy can be evaluated.
These findings are at variance with traditional claims that the gradient
corrections vanish in all orders.
Having concluded that the derivations that support these claims are
problematic in the transition from the joint density-potential functional to
the density-only functional, we recommend working consistently with the joint
functional.

\heading{Acknowledgments}
We thank P. Trevisanutto for valuable discussions.
This work is funded by the Singapore Ministry of Education
and the National Research Foundation of Singapore.  
H.~K.~N is also funded by a Yale-NUS College start-up grant.
C.~A.~M. acknowledges the hospitality of Institut Non Lin\'eaire de Nice (CNRS
and Universit\'e de Nice) and Laboratoire de Physique Th\'eorique
(CNRS and UPS Toulouse).


\begin{thebibliography}{10}
\bibitem{rp-RP} 
We use lower case letters $\vec{r}$, $\vec{p}$ for the classical
phase space variables and upper case letters $\vec{R}$, $\vec{P}$ for 
the operators.
\bibitem{ourEX}
The example of Eqs.~(\ref{eq:A12})--(\ref{eq:A18}) is a bit pathological in
this respect, but an easy cure is available:
Replace the contact interaction with an interaction that has a short
range $a>0$,\newline
\centerline{$\displaystyle%
  W\delta(\vec{r})\to W\int\frac{(\D\vec{k})}{(2\pi)^2}
  \frac{\Exp{\I\vec{k}\cdot\vec{r}}}{1+a^2\vec{k}^2}
  =\frac{W}{2\pi a^2}K_0\biggl(\frac{|\vec{r}|}{a}\biggr)\,.\hspace*{2em}$}
Solving the corresponding version of \Eq{A14b} for $n(\vec{r})$ is then
accomplished by
$n(\vec{r})=\frac{1}{W}\bigl(1-a^2\grad^2\bigr)
             \bigl[V(\vec{r})-V_{\mathrm{ext}}(\vec{r})\bigr]$.
With $n(\vec{r})$ from \Eq{A14a}, this is a second-order differential equation
for $V(\vec{r})$, and the resulting effective potential energy has a continuous
gradient field.
\bibitem{vW}
The correction term $\propto\delta(n)(\grad n)^2$ differs in structure from
the well-known vW correction $\propto n^{-1}(\grad n)^2$ to the kinetic energy
functional in 3D, and the analogous 1D term (which, however, has a negative
coefficient).
\bibitem{KKvsW} 
The Kirkwood and Kirzhnits (K\&K) expansion methods are particularly popular. 
They are similar to, but different from, the Wigner-function based methods.   
In the K\&K approaches, operator functions of $\vec{R}$ and $\vec{P}$ are
ordered such that all $\vec{R}$s are to the left of all $\vec{P}$s in products
(or vice versa), whereas a symmetric ordering underlies
the Wigner-function methods.
\bibitem{V-func}
There is, in particular, Schwinger's derivation of the leading and second
corrections to the TF energy of neutral atoms
\cite{Schwinger:80,Schwinger:81}; see also \cite{Englert:88} and references
therein.
\bibitem{KSvsSemiclassics}
For a meaningful comparison of the semiclassical approximation in \Eq{B4} with the exact (KS) evaluation of the trace in \Eq{B1}, one should use the same effective potential energy in both expressions. Therefore, the use of noninteracting fermions is expedient in this context since the two self-consistent solutions of Eqs.~\eqref{eq:A10} --- one with the $E_1[V-\mu]$ of \Eq{B1}, the other with that of \Eq{B4} --- yield different $V(\vec r)$s if ${E_{\mathrm{int}}[n]\neq0}$.
\bibitem{HarmOscEnergy}
The exact energy is
\newline\centerline{$\displaystyle%
\frac{E(N)}{\hbar\omega}=\tfrac{2}{3}N
{\left(N+\tfrac{1}{4}\right)}^{\frac{1}{2}}
+{\left(\tfrac{1}{4}-y^2\right)}{\left[
{\left(N+\tfrac{1}{4}\right)}^{\frac{1}{2}}-\tfrac{1}{3}y\right]}
\,,\hspace*{2.5em}$}
where $y(N)$ is the difference between
${\left(N+\tfrac{1}{4}\right)}^{\frac{1}{2}}$ and its nearest integer,
with ${y=\pm\frac{1}{2}}$ 
for closed shells.
\bibitem{Airy-av} 
One main ingredient is the Airy-averaged version of \Eq{B2},
\newline\centerline{$\displaystyle%
[g(A)]_{\textsc{w}}=\int\D x\,\mathrm{Ai}(x)\Bigl[g(a_x)
   -\frac{\hbar^2}{16}\bigl\{A_\textsc{w}\Lambda^2A_\textsc{w}\bigr\}
    g''(a_x)\Bigr]\,,\hspace*{2em}$}
where ${a_x=A_{\textsc{w}}+\frac{x}{2}\bigl(\hbar^2\bigl\{A_\textsc{w}\Lambda
  A_\textsc{w}\Lambda A_\textsc{w}\bigr\}\bigr)^{1/3}}$ and
$\mathrm{Ai}(\ )$  is the Airy function.
The right-hand side agrees with that of \Eq{B2} to order $(\hbar\Lambda)^2$
and includes contributions from higher-order terms in a systematic way.
It gives the exact $[g(A)]_{\textsc{w}}$ for
${A=H-\mu=\frac{1}{2m}\vec{P}^2+V(\vec{R})-\mu}$ if the effective potential
energy is that of a constant force.
\bibitem{Alt-E1}
If there are 2D and 3D analogs of the 1D densities of \cite{Ribeiro+4:15},
they offer another strategy for constructing improved approximations for
$E_1[V-\mu]$.
\end{thebibliography}
\end{document}